\documentclass[lettersize,journal,twoside]{IEEEtran}

\IEEEoverridecommandlockouts                              

\usepackage{amssymb}  
\usepackage{dsfont}	
\usepackage{amsfonts,amssymb}
\usepackage{multirow}
\usepackage{color}
\usepackage{subfigure}
\usepackage{soul}
\usepackage{ulem}
\usepackage{balance}
\usepackage{xcolor}  

\usepackage[letterpaper,top=60pt,bottom=48pt,left=48pt,right=43pt]{geometry}
\usepackage{hyperref}
\hypersetup{
	colorlinks=true,
	linkcolor=blue,
	citecolor=blue,
	urlcolor=black
}

\title{sEMG-based Joint Angle Estimation via Hierarchical Spiking Attentional Feature Decomposition Network}

\author{Xin Zhou$^{1}$, Chuang Lin$^{2}$, Can Wang$^{3}$ and Xiaojiang Peng$^{1,*}$ 
\thanks{Manuscript received: September 10, 2024; Revised: November 26, 2024; Accepted: December 25, 2024.}
\thanks{This paper was recommended for publication by Ryu, Jee-Hwan upon evaluation of the Associate Editor and Reviewers’ comments.}
\thanks{This work is partially supported by the National Natural Science Foundation of China (62176165), the
	Stable Support Projects for Shenzhen Higher Education Institutions (20220718110918001), the Natural
	Science Foundation of Top Talent of SZTU (GDRC202131), the Basic and Applied Basic Research
	Project of Guangdong Province (2022B1515130009), and the Special subject on Agriculture and Social
	Development, Key Research and Development Plan in Guangzhou (2023B03J0172).}
\thanks{$^1$ College of Big Data and Internet, Shenzhen Technology University, Shenzhen, 518118, China. (zhouxin, pengxiaojiang)@sztu.edu.cn}
\thanks{$^2$ School of Information Science and Technology, Dalian Maritime University, Dalian 116026, China. linchuang\_78@126.com}
\thanks{$^3$ Shenzhen Institutes of Advanced Technology, Chinese Academy of Sciences, Shenzhen 518055, China. can.wang@siat.ac.cn}
\thanks{$^*$ Corresponding author.}
}

\newcommand{\RNum}[1]{\uppercase\expandafter{\romannumeral #1\relax}}
\usepackage{graphicx} 
\usepackage{amsmath}
\usepackage{booktabs}
\usepackage{threeparttable}
\usepackage{cite}
\UseRawInputEncoding  
\usepackage[letterpaper,top=60pt,bottom=48pt,left=48pt,right=43pt]{geometry}
\begin{document}

\maketitle

\begin{abstract}

Surface electromyography (sEMG) has demonstrated significant potential in simultaneous and proportional control (SPC). However, existing algorithms for predicting joint angles based on sEMG often suffer from high inference costs or are limited to specific subjects rather than multi-subject scenarios. To address these challenges, we introduced a hierarchical Spiking Attentional Feature Decomposition Network (SAFE-Net). This network initially compresses sEMG signals into neural spiking forms using a Spiking Sparse Attention Encoder (SSAE). Subsequently, the compressed features are decomposed into kinematic and biological features through a Spiking Attentional Feature Decomposition (SAFD) module. Finally, the kinematic and biological features are used to predict joint angles and identify subject identities, respectively. Our validation on two datasets and comparison with two existing methods, Informer and Spikformer, demonstrate that SSAE achieves significant power consumption savings of 39.1\% and 37.5\% respectively over them in terms of inference costs. Furthermore, SAFE-Net surpasses Informer and Spikformer in recognition accuracy on both datasets. This study underscores the potential of SAFE-Net to advance the field of SPC in lower limb rehabilitation exoskeleton robots. The code and model for SAFE-Net are available at \href{https://github.com/xzhoubu/SAFE\_Net}{https://github.com/xzhoubu/SAFE\_Net}.



\end{abstract}

\section{Introduction}
\label{sec:introduction}
\IEEEPARstart{I}{n} recent years, an increasing number of individuals suffer from neurological impairments, muscle atrophy, or restricted mobility, such as stroke survivors, spinal cord injury patients, and amputees. This phenomenon has led to a growing demand for rehabilitation exoskeleton robots. Providing continuous real-time intention detection strategies for human-machine interaction systems is one of the key steps towards achieving human-machine collaboration. Therefore, the existence of an efficient, real-time, and robust motion intention detection strategy would significantly enhance the coordination and continuity of rehabilitation exoskeleton robots.

Surface electromyography (sEMG) signals are widely used in motion intention recognition tasks due to their stability and non-invasiveness, including gesture recognition, behavioral pattern recognition and gait phase recognition \cite{tigrini2024phasor}. These discrete motion intention detection tasks typically serve applications such as prosthetic control, gait phase anomaly detection, and fall monitoring. However, discrete motion intention recognition algorithms can only identify limited behavioral patterns and cannot meet the requirements of simultaneous and proportional control (SPC) for continuity and real-time performance. sEMG-based continuous lower-limb joint angle estimation is pivotal in rehabilitation robotics, wearable exoskeletons, and assistive devices \cite{mobarak2024minimal}. These applications require precise and real-time joint angle predictions to enable seamless human-machine interaction and achieve simultaneous and proportional control (SPC), enhancing mobility and quality of life for users.

SPC algorithms for continuous motion intention recognition typically focus on estimating joint angles, such as shoulder-elbow joints \cite{bi2019review}, finger joints \cite{ma2021novel}, and hip-knee-ankle joints \cite{wang2022prediction}. These methods typically collect kinematic, dynamic, or physiological signals as inputs to the model and then use artificial neural networks to predict the required joint angle values. Recent studies have explored the application of Transformer-based architectures for sEMG signal processing. Liang et al. \cite{liang2023semg} proposed a Tightly Coupled Convolutional Transformer (TCCT) model for knee joint angle estimation, demonstrating superior performance in prediction accuracy and computational efficiency compared to traditional methods. Similarly, SCTNet proposed by An et al. \cite{an2024sctnet} integrated shifted windows and convolutional layers, effectively capturing both local and global sEMG features, achieving competitive results in fine-grained finger joint angle estimation. Additionally, a BERT-inspired framework was developed by Lin et al.  \cite{lin2022bert}, focusing on cross-subject hand kinematics estimation, which emphasized robustness and transferability across diverse subjects. These end-to-end prediction modes require less domain knowledge while achieving decent prediction accuracy. Still, little attention has been paid to the inference latency and power consumption required to use these deep neural networks. Computational efficiency typically relates to the flexibility and coordination of human-machine interaction systems.

Utilizing bio-inspired algorithms to optimize artificial neural networks and accelerate model convergence is a feasible strategy for improving computational efficiency, such as using genetic algorithms to optimize the weights and thresholds of neurons. Some studies have employed particle swarm optimization to optimize back propagation neural networks \cite{jin2021long} and long short-term memory (LSTM) networks \cite{sohane2022single}. Recently, spiking neural networks (SNNs) have garnered attention in computational science due to their sparse computation, event-driven nature, and ultra-low power consumption \cite{fang2023spikingjelly, wu2021tandem, wu2021progressive}. Moreover, SNNs simulate the working mechanism of human brain neurons for computation, making them more biologically plausible. SNNs encode continuous biological signals into discrete spiking neural signals. Essentially, SNNs do not transmit complete continuous biological signal values but rather release a spiking signal when the membrane potential reaches a threshold. Ultimately, binary encoding is transmitted within the network. This significantly reduces storage costs and computational complexity, and also provides possibilities for optimizing SPC algorithms.

Benefiting from the spike-based event-driven computational characteristics of binary spiking signals, sequences encoded using SNNs require less complexity in multiplication compared to multiplying two floating-point numbers, but only need to sum the corresponding non-zero positions. Specifically, incorporating the spike-driven computational paradigm into attention mechanisms allows for sparse addition operations throughout the entire network \cite{yao2024spike, zhou2022spikformer}. Further reduction in power consumption is achieved by pruning input spikes using random masking techniques \cite{wang2023masked}. The utilization of biologically plausible spiking neurons and event-driven neural encoders enables the processing of electroencephalogram signals, demonstrating the potential of SNNs in decoding biological signals \cite{cai2023bio, faghihi2022neuroscience}. Additionally, Tieck et al. \cite{tieck2020spiking} used sEMG and SNN methods to control robotic arms to understand human motor control and sensory processing mechanisms, and applied them to robotics. These studies open new possibilities for developing human-machine interaction systems.

Although sEMG contains a wealth of motion information, its heterogeneity poses a challenge to the generalization ability of a single model. That is, the sEMG signals collected from the same individual in different sessions or from different individuals performing the same action in the same session are varied.  This challenges the generalization ability of a single model. Utilizing transfer learning strategies to improve gesture recognition systems and alleviate training burdens is feasible \cite{chen2020hand}. For example, Wang et al. \cite{wang2023similarity} constructed Siamese neural networks, enabling rapid deployment of classifiers applicable to new scenarios with only a small number of samples per class. Additionally, incorporating prior knowledge \cite{zhou2023continuous}, employing unsupervised strategies for data augmentation \cite{li2023cross}, or constructing graph topology structures \cite{xu2023cross} can aid in multi-subject electromyography pattern recognition (MPR). However, these methods rely on manually constructed external knowledge. Similarly, few-shot learning contributes to domain generalization problems, where learning mappings using only a small amount of data 
and quickly adapting to new gestures through prior experience \cite{rahimian2021few, rahimian2021fs}. Although few-shot learning requires few manual annotations, it may also lead to overfitting issues.

Inspired by computer vision tasks such as facial expression recognition \cite{zou2022facial, ruan2021feature}, a compound expression can be composed of domain-specific and domain-agnostic features weighted together. More specifically, a happy expression can be a combination of a basic action like smiling with subtle variations \cite{zou2022learn}. We believe that human walking can also be decomposed into kinematic features and biological features. Kinematic features ignore individual differences and focus on kinematic information, while biological features are more concerned with individual identity differences. 

To this end, we design a hierarchical Spiking Attentional FEature decomposition Network (SAFE-Net) which first compresses sEMG signals into coarse-grained features using a Spike-driven Sparse Attention Encoder (SSAE). Then, the compressed features are hierarchically decomposed into fine-grained kinematic features and biological features using a Spiking Attentional Feature Decomposition (SAFD) module and finally regressing kinematic features into joint angle information. The biological information can be used for the subject identity recognition. The main contributions of this study are as follows:

\begin{itemize}
	
	\item We developed a Spike-driven Sparse Attention Encoder (SSAE) for compressing sEMG features, which not only improves the accuracy of MPR but also significantly enhances inference efficiency with lower power consumption.
	
	\item We proposed a hierarchical Spiking Attentional Feature Decomposition (SAFD) strategy, which can decompose sEMG features into fine-grained kinematic features and biological features, aiming to improve the accuracy of joint angle recognition in multi-subject scenarios.
	
	\item Experiments were conducted to confirm that SSAE and SAFE-Net achieve higher recognition accuracy on two different motion pattern datasets compared to two Transformer-based methods, Informer and Spikformer.
	
\end{itemize}

The remaining sections of this article are organized as follows. The Section \ref{sec2} presents the SAFE-Net pipeline for decomposing sEMG signals. The Section \ref{sec3} details the procedure of the conducted experiments. Experimental results are presented in the Section \ref{sec4}. A more in-depth discussion is provided in the Section \ref{sec5}. Finally, the study is concluded in the Section \ref{sec6}.

\begin{figure*}[thpb]
	\centering
	\includegraphics[width=0.87\linewidth]{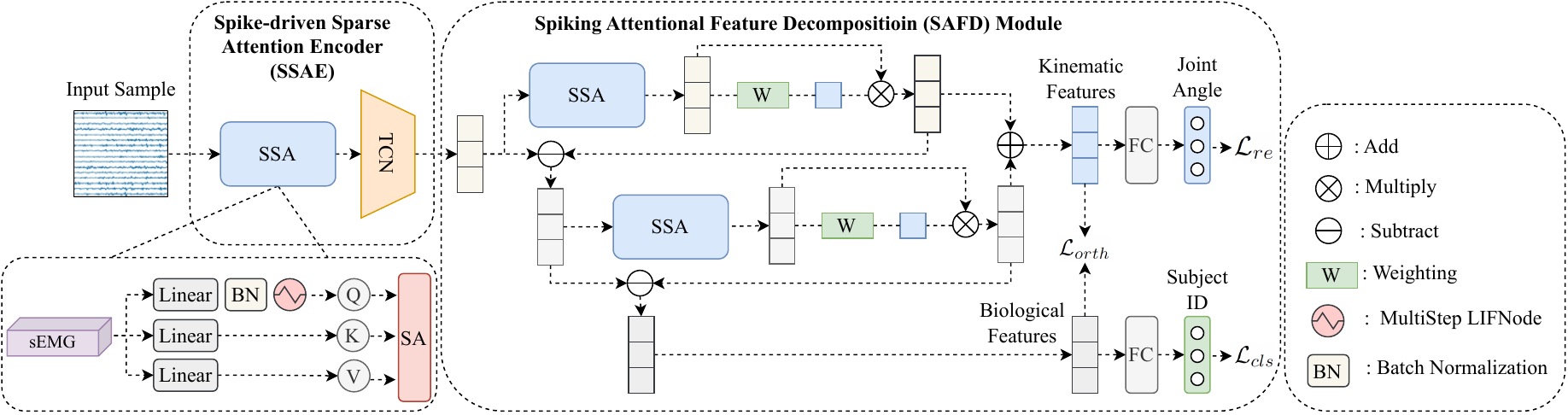}
	\caption{The overall schematic of the proposed SAFE-Net consists of two parts: a Spike-driven Sparse Attention Encoder (SSAE) and a Spiking Attentional Feature Decomposition (SAFD) module. Taking sEMG signals as input, SAFE-Net is trained to estimate lower limb joint angles and subject identity recognition. SSA: Spike-driven Sparse Attention. SA: Sparse Attention. TCN: Temporal Convolution Network. FC: Fully-Connected layer.}
	\label{fig.safe-net}
\end{figure*}

\section{Methods}\label{sec2}

In this section, we formalize the SAFE-Net for decomposing sEMG signals. As shown in Fig. \ref{fig.safe-net}, the network consists of two main modules: the SSAE and the SAFD module. We will discuss each module in detail below.

\begin{figure}[htbp]
	\centering{
		\subfigure[{Raw sEMG signal.}]{
			\includegraphics[height=3cm,width=0.43\linewidth]{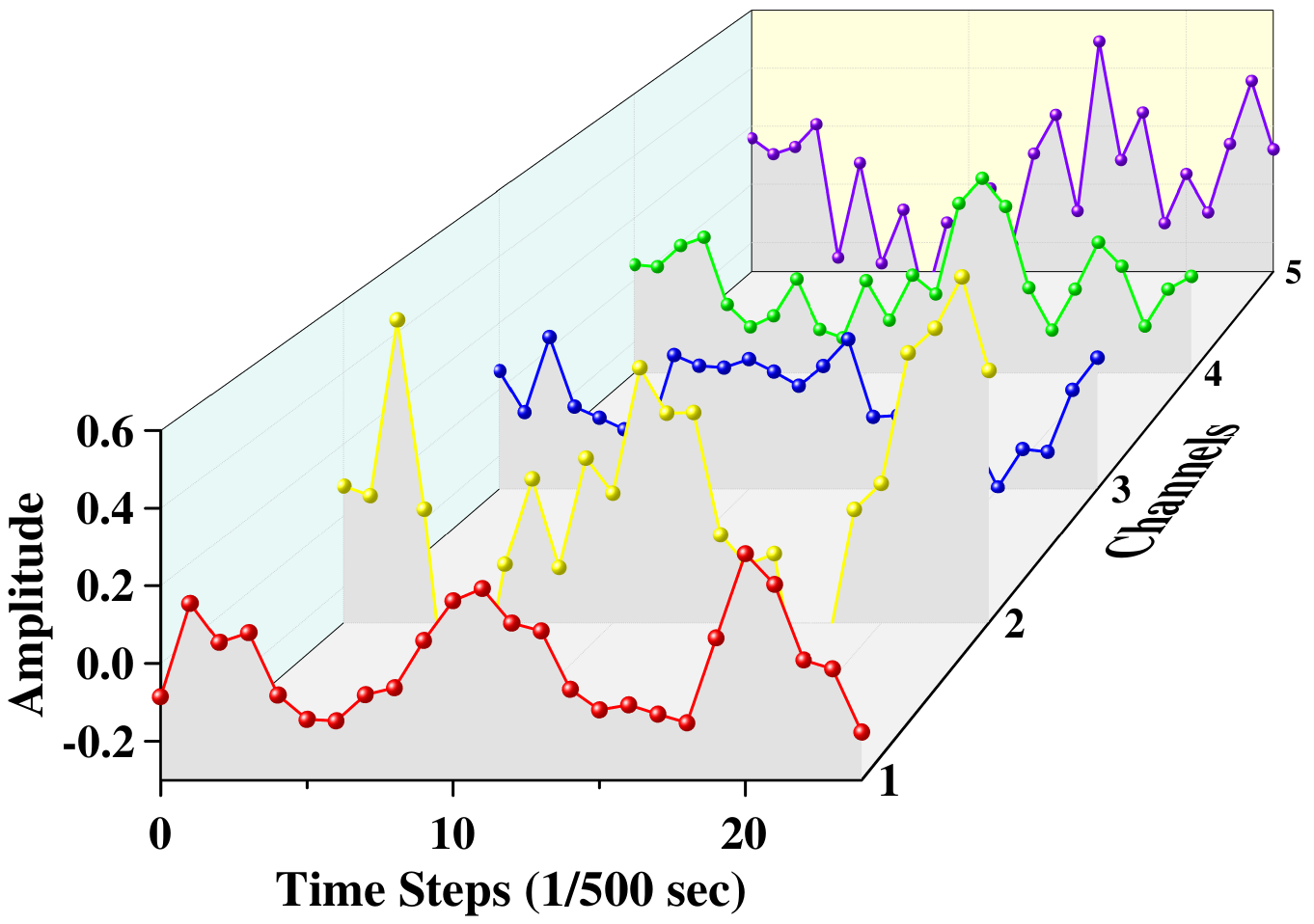}
			\label{fig.raw}
		}
		\quad
		\subfigure[{Neural spiking signal.}]{
			\includegraphics[height=3cm,width=0.42\linewidth]{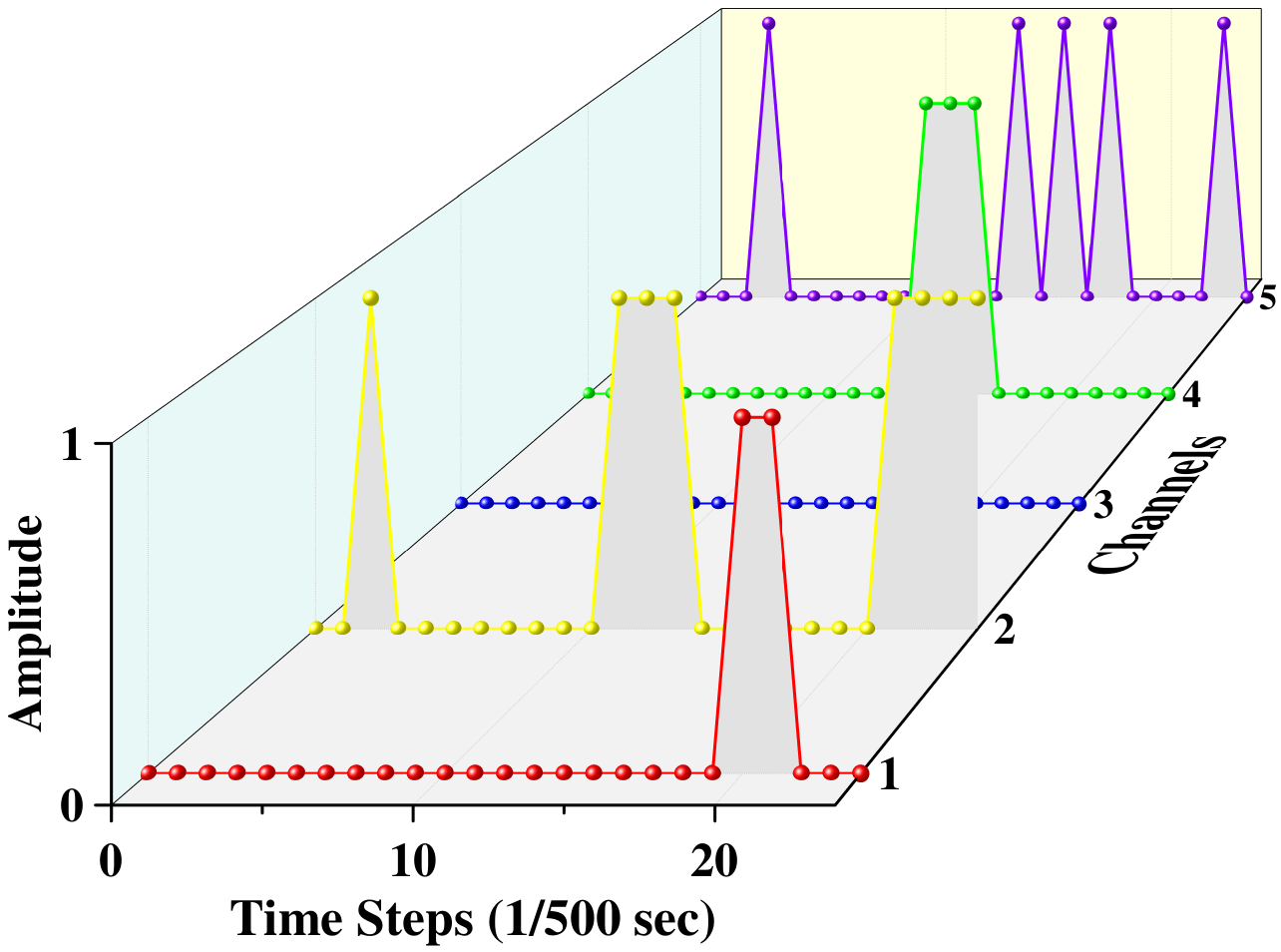}
			\label{fig.spiking}
		}
		
	}
	
	\caption{
		Visualizations of the raw sEMG signal and neural spiking signal after compressing by a LIFNode.}
	\label{fig.signal}
\end{figure}

\subsection{Spike-driven Sparse Attention Encoder (SSAE)}

\subsubsection{Spiking neurons}
Spiking neurons are the fundamental units of SNNs. Besides spatial information propagation, SNNs also consider temporal historical information's impact on the network's current state. Additionally, SNNs are event-driven computational models that arouses spiking signals only when the membrane potential exceeds a threshold, which activates subsequent neurons and facilitates information transmission. Fig. \ref{fig.signal} visually illustrates how spiking neurons compress the five-channel sEMG signals into neural spiking signals. This operational mechanism makes SNNs more energy-efficient. Furthermore, SNNs are closer to biological neuron mechanisms, and therefore exhibits biological interpretability.

The operation of spiking neurons can be summarized into three stages: 1) Input signal reception (charging): spiking neurons receive signals from other units through dendrites. 2) Integration and spiking emission (discharging): spiking neurons integrate received input signals in the soma. When the integration value surpasses the neuron's threshold, a spiking neuron generates an action potential. 3) Recovery and reset (resetting): After the generation of the spike, spiking neurons return to their base state and can integrate input signals again. Specifically, the charging process of a Leaky Integrate-and-Fire (LIF) neuron can be formalized as the following equation:

\begin{equation}
	\label{eq.1}
	\tau\frac{d H(t)}{d t}=-\left(H[t]-H_{rest }\right)+X(t),
\end{equation}

\noindent where $\tau$ represents the membrane potential constant, $H[t]$ and $H_{rest}$ represent the membrane potential at time $t$ and $rest$, respectively. $X[t]$ denotes the current input. At each time $t$, the change of membrane potential satisfies Equation \ref{eq.1}. $H[t]$ should fall ($H[t] - H_{rest}$), and at the same time, $H[t]$ should rise by a value which is contributed by $X[t]$. The discharge process is shown in Equation \ref{eq.2}:

\begin{equation}
	\label{eq.2}
	\left\{\begin{array}{c}
		S(t)=1, \text { if } H(t) \geq H_{t h} \\
		S(t)=0, \text { if } H(t) < H_{th}.
	\end{array}\right.
\end{equation}

\noindent When the membrane potential $H[t]$ accumulates to the threshold value $H_{th}$, the spiking neuron performs a discharge operation. At this point, the output membrane potential $S[t]$ is set to 1, and $H[t]$ is reset to the reset potential $H_{rest}$. When $H[t]$ is below the threshold $H_{th}$, $S[t]$ remains 0 at the time. After the neuron completes a discharge operation, it is reset. Typically, there are two methods for resetting:

\begin{equation}
	\lim _{\Delta t \rightarrow 0^{+}} H(t+\Delta t)=\left\{\begin{array}{c}
		H_{reset}, \text { hard reset } \\
		H(t)-H_{th}, \text { soft reset }.
	\end{array}\right.
\end{equation}

\noindent In this study, we only employ hard reset method because in our work, we directly train SNNs, which yields better experimental performance \cite{fang2023spikingjelly}. Since discharge operation is triggered only when the membrane potential exceeds a threshold value, the entire spiking sequence consists of sparse binary spiking values.

\subsubsection{Spike-driven Sparse Attention (SSA)}
Given a sEMG signal record $I$ = ($i_1$, $i_2$, $\cdots$, $i_{t}$), $I$ $\in$  $\mathbb{R}^{t\times c}$, and a target joint angle record $Y$ = ($y_1$, $y_2$, $\cdots$, $y_{t}$), $Y$ $\in$  $\mathbb{R}^{t\times n}$, where $t$ represents the time steps, and $c$ and $n$ represent the number of channels collected by sEMG sensors and the number of target angle channels to be predicted, respectively. The signal $I$ first passes through the data embedding layer, which includes value embedding and positional embedding. The value embedding layer is implemented by one-dimensional convolution, while the positional embedding layer adopts sine and cosine encoding. The results of the two parts are summed to obtain the dimensionally expanded embedded data $E$ = ($e_1$, $e_2$, $\cdots$, $e_{t}$), $E$ $\in$  $\mathbb{R}^{t\times d}$, where $d$ represents the number of hidden layer features. Then, $E$ is mapped into matrices $\mathbf{\textit{Q}}$, $\mathbf{\textit{K}}$, and $\mathbf{\textit{V}}$ in different ways as follows:

\begin{equation}\label{eq:QKV}
	\mathbf{\textit{Q}}=SN(BN({EW}^{\mathbf{\textit{Q}}})),\mathbf{\textit{K}}={EW}^{\mathbf{\textit{K}}},{\mathbf{\textit{V}}}={EW}^{\mathbf{\textit{V}}},
\end{equation}
\noindent where ${W}^{\mathbf{\textit{Q}}}$, ${W}^{\mathbf{\textit{K}}}$, ${W}^{\mathbf{\textit{V}}}$ $\in$ $\mathbb{R}^{d\times d}$ are the $query$, $key$ and $value$ projection matrices, respectively. $BN(\cdot)$ and $SN(\cdot)$ denote batch normalization and spiking neuron layer, respectively.

Typically, the space-time complexity of computing a self-attention map for a time series of length $L$ is $L^2$. This often requires powerful computility to achieve low latency. To this end, we filter out the $\mathbf{\textit{Q}}$, $\mathbf{\textit{K}}$ dot product pairs and for those close to the uniform distribution (referred to as Lazy $query$), we fill the attention map with the mean value. For dot product pairs far from a uniform distribution (referred to as active $query$, $\overline{\mathbf{\textit{Q}}}$), we retain the original calculated values. In this study, we utilize the Kullback-Leibler divergence \cite{zhou2021informer} to assess what extent a probability distribution approximates a uniform distribution. The specific calculation method for sparse attention is as follows:

\begin{equation}
	\label{sparse attention}
	\mathcal{A}(\overline{\mathbf{\textit{Q}}}, \mathbf{\textit{K}}, \mathbf{\textit{V}})=\operatorname{Softmax}\left(\frac{\overline{\mathbf{\textit{Q}}} \mathbf{\textit{K}}^{T}}{\sqrt{d}}\right) \mathbf{\textit{V}}. \\
\end{equation}

\begin{figure}[t]
	\centering
	\includegraphics[width=0.8\linewidth]{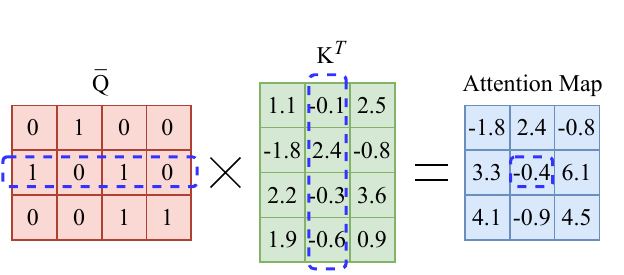}
	\caption{
		Illustration to the mechanism of spiking sparse attention. The values of the elements in the Attention Map are equal to the values summed over the positions in the $\mathbf{\textit{K}}^{T}$ matrix corresponding to the elements of the  non-zero rows of $\overline{\mathbf{\textit{Q}}}$. In this example, the non-zero row elements in $\overline{\mathbf{\textit{Q}}}$ are located at the first and third positions, and the corresponding position values in $\mathbf{\textit{K}}^{T}$ are -0.1 and -0.3, respectively. Thus, -0.4 = -0.1 + (-0.3).}
	\label{fig.SSA}
\end{figure}

It is worth noting that we only pass through a spike neuron when obtaining the $\overline{\mathbf{\textit{Q}}}$ matrix, and we choose linear mapping directly when obtaining the $\mathbf{\textit{K}}$ and $\mathbf{\textit{V}}$ matrices. When computing $\overline{\mathbf{\textit{Q}}} \mathbf{\textit{K}}^{T}$, since $\overline{\mathbf{\textit{Q}}}$ is a binary matrix with elements of 0 or 1, the dot product operation is equivalent to summing the values of the corresponding positions of the elements of the non-zero rows of $\overline{\mathbf{\textit{Q}}}$ in the $\mathbf{\textit{K}}$ matrix. As shown in Fig. \ref{fig.SSA}, the selected row vectors in $\overline{\mathbf{\textit{Q}}}$ multiplied by the column vectors in $\mathbf{\textit{K}}^{T}$ are equivalent to summing the values at the first and third positions in $\mathbf{\textit{K}}^{T}$. Whether $\overline{\mathbf{\textit{Q}}}$ or $\mathbf{\textit{K}}^{T}$ is transformed into a spiking binary matrix will force the dot product operation to become a summation operation. Therefore, in order to maintain a balance between preserving information and not sacrificing prediction accuracy, we only use spike neurons when obtaining $\overline{\mathbf{\textit{Q}}}$.

\subsubsection{Temporal convolution network (TCN)}
After the calculation of spiking sparse attention, we utilize a TCN layer to establish temporal dependencies. TCN mainly consists of two mechanisms: causal convolution and dilated convolution. Causal convolution ensures that each position of the output depends only on the historical data, while dilated convolution can expand the receptive field to capture long-range dependencies. Causal dilated convolution can simultaneously perceive input information at different distances through convolution kernels with different dilation rates. This means that within a single layer of the network, a nonlinear receptive field can be established by causal dilated convolution, while keeping the number of layers logarithmically related to the length of the input sequence. This improve the expressive power of the network \cite{zhou2023continuous}.

\subsection{Spiking Attentional Feature Decomposition (SAFD)}
As shown in Fig. \ref{fig.safe-net}, SAFD consists of several SSA and weighting modules, based on a cascaded feature decomposition mechanism to obtain kinematic and biological features. Specifically, for input feature $x_1$, where $x_1$ $\in$ $\mathbb{R}^{d}$, it first passes through $SSA$ to obtain the base feature $p_1$, which is represented by the following formula:

\begin{equation}
	p_{1} = SSA(x_{1}).
\end{equation}

\noindent Then, $p_{1}$ undergoes weighting calculation using the module $W(\cdot)$ to obtain the corresponding weight $w_1$, expressed as follows:

\begin{equation}
	w_{1} = W(p_{1}) = W(SSA(x_{1})).
\end{equation}

\noindent The meta-kinematic and meta-biological feature are obtained by the following mathematical forms:

\begin{equation}
	q_{1} = w_{1} \cdot p_{1},
\end{equation}
\begin{equation}
	r_{1} = x_{1} - q_{1}.
\end{equation}

\noindent After $i$ iterations of SAFD module, we obtain meta-kinematic feature sets ($q_1$, $q_2$, $\cdots$, $q_{i}$) and meta-biological feature sets ($r_1$, $r_2$, $\cdots$, $r_{i}$). The final kinematic feature $F_k$ is calculated as the sum of $q_i$, and the biological feature $F_b$ is equal to $r_i$. $F_k$ and $F_b$ can be used for joint angle estimation and subject identity recognition respectively.

\subsection{Optimization and Training}
SAFE-Net takes a window-sized sEMG signal as input and outputs the corresponding joint angle values and the identity ID of the subject. The network updates its parameters using the back propagation algorithm. For each iteration, SAFE-Net randomly selects a batch of data ($X$, $Y_r$, $Y_c$), where $X$, $Y_r$, $Y_c$ represent the sEMG signal, the corresponding joint angles, and the subject identity labels, respectively. $X$ is fed into SAFD-Net to obtain kinematic feature $F_k$ and biological feature $F_b$. On the one hand, $F_k$ is used to predict joint angles, and the regression loss $\mathcal{L}_{re}$ is defined by the mean squared error (MSE):

\begin{equation}
	\operatorname{MSE}=\frac{1}{\mathrm{~N}} \sum_{\mathrm{i}=1}^{\mathrm{n}}\left(\overline{\mathrm{Y}^{i}_{\mathrm{r}}}-\mathrm{Y}^{i}_{\mathrm{r}}\right)^2,
\end{equation}

\noindent where $\overline{\mathrm{Y}^{i}_{\mathrm{r}}}$ and $\mathrm{Y}^{i}_{\mathrm{r}}$ represent the predicted angle and the true angle respectively. On the other hand, $F_b$ is used for subject identity prediction, and the classification loss $\mathcal{L}_{cls}$ is defined by the cross-entropy loss:

\begin{equation}
	\mathcal{L}_{c l s}^p=-\sum_{c=1}^{C} \mathds{1}_{\left[c=y_c\right]} \log \left(f\left(F_b\right)\right),
\end{equation}

\noindent where $f(\cdot)$ is the linear classifier, $C$ is the number of categories, and $\mathds{1}_{[c=y_c]}$ equals 1 when the predicted category matches the true category, otherwise it equals 0. To prevent overfitting in the identity recognition task, we introduce an orthogonal regularization strategy. Specifically, we calculate the orthogonal loss for $F_k$ and $F_b$ as follows:

\begin{equation}
	\mathcal{L}_{orth}= F_k \cdot F_b.
\end{equation}

\noindent Therefore, the total loss is calculated as follows:

\begin{equation}
	\mathcal{L}_{total}= \alpha \mathcal{L}_{re} + \beta \mathcal{L}_{cls} + \gamma \mathcal{L}_{orth},
\end{equation}

\noindent where $\alpha$, $\beta$, and $\gamma$ are balancing weight coefficients. $\mathcal{L}_{cls}$ penalizes samples that are misclassified in terms of subject identity, while $\mathcal{L}_{orth}$ ensures that the kinematic and biological features are orthogonal.


\begin{figure}[t]
	\centering
	\includegraphics[width=0.85\linewidth]{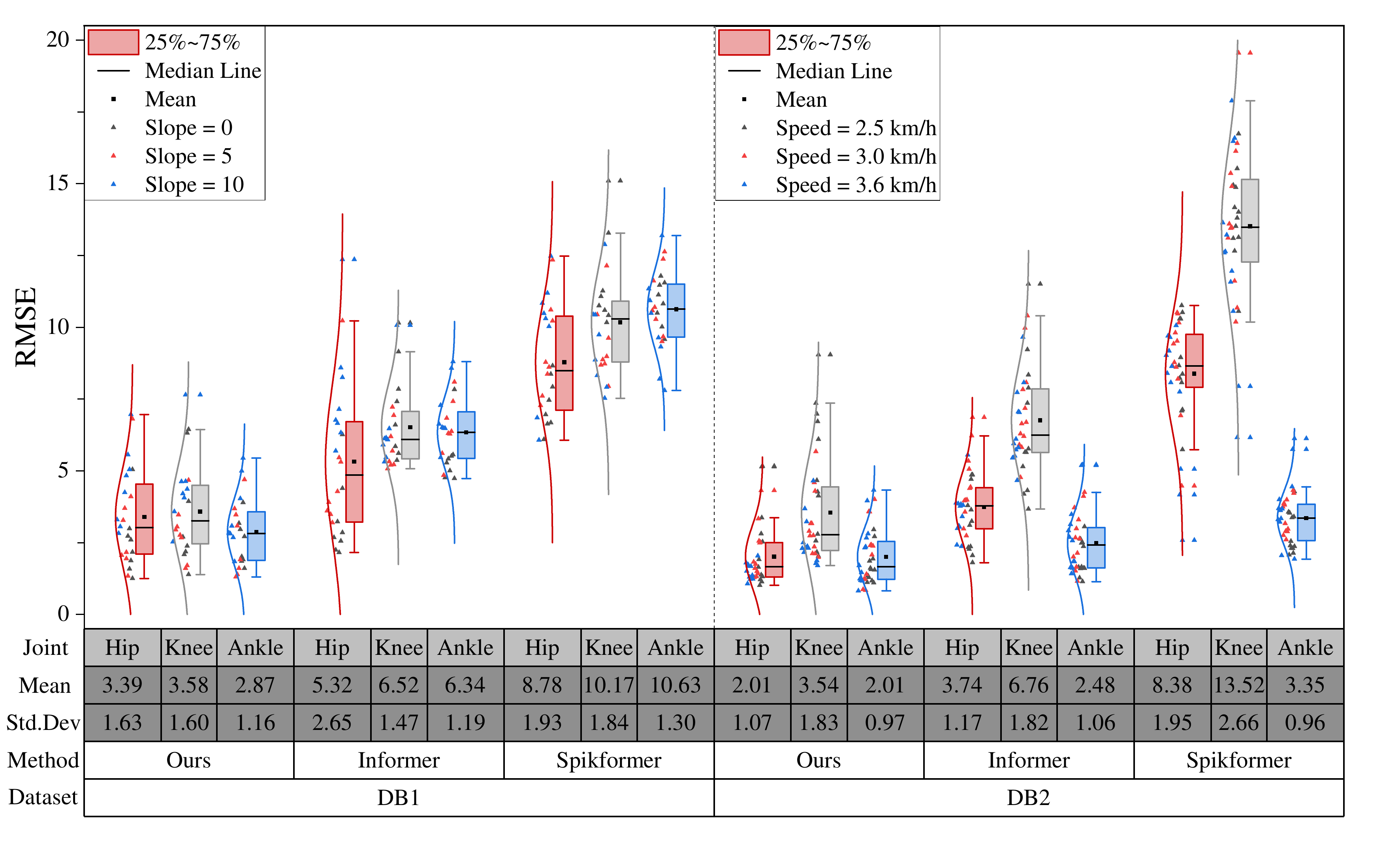}
	\caption{The box plots compare the RMSE of hip-knee-ankle joint prediction results among SAFE-Net, Informer, and Spikformer on datasets DB1 and DB2. Each box represents the RMSE of different methods, where a lower box height indicates smaller errors. Each small triangle represents the RMSE of an individual, with denser clustering indicating lower variance in prediction accuracy for that method.}
	\label{fig.diff_methods}
\end{figure}

\begin{table}[htbp]
	\centering
	\caption{Hyperparameter Settings}
	\begin{tabular}{ccc}
		\toprule
		Hyperparameter & Description  & Value \\
		\hline
		{$d$} &Hidden layer feature number   &64  \\ 
		{$\tau$} &Membrane time constant     &2  \\ 
		{$H_{th}$} &Threshold voltage of neurons   &0.3  \\ 
		{$\alpha$} &Regression loss coefficient     &0.1  \\
		{$\beta$} &Classification loss coefficient     &1  \\
		{$\gamma$} &Orthogonal  loss coefficient     &0.5  \\
		\bottomrule
	\end{tabular}
	\label{tab1}
\end{table}

\begin{table*}[h]
	\caption{
		Comparison of RMSE, MAE, PCC, and R\textsuperscript{2} between different models on DB1 and DB2.}
	\resizebox{\textwidth}{!}{
		\begin{tabular}{c|c|cccc|cccc|cccc}
			\hline \hline
			\multicolumn{1}{c|}{\multirow{8}{*}{DB1}}                                    &
			\multicolumn{1}{c|}{\multirow{2}{*}{Method}}                                                                      & 
\multicolumn{4}{c|}{Slope = 0}  & \multicolumn{4}{c|}{Slope = 5}   &\multicolumn{4}{c}{Slope = 10}
			\\ \cline{3-14}
			\multicolumn{1}{c|}{}      &\multicolumn{1}{c|}{}                                     
            & \multicolumn{1}{c}{RMSE}          & \multicolumn{1}{c}{MAE}  &\multicolumn{1}{c}{PCC}             &\multicolumn{1}{c|}{R\textsuperscript{2}}        
			& \multicolumn{1}{c}{RMSE}               & \multicolumn{1}{c}{MAE}            & \multicolumn{1}{c}{PCC}              &\multicolumn{1}{c|}{R\textsuperscript{2}}
			& \multicolumn{1}{c}{RMSE}               & \multicolumn{1}{c}{MAE}            & \multicolumn{1}{c}{PCC}               &\multicolumn{1}{c}{R\textsuperscript{2}}      \\ \cline{2-14}
			
			\multicolumn{1}{c|}{}                   & SSAE  
			&\textbf{4.015}	&\textbf{3.069}	&\textbf{0.970} &\textbf{0.937}	&\textbf{4.278}	&\textbf{3.335}	&{0.965} &\textbf{0.928}	&\textbf{5.793}	&\textbf{4.589}	&\textbf{0.943} &\textbf{0.869}
			\\ 
			\multicolumn{1}{c|}{}                                              & Informer \cite{zhou2021informer}                 
			& 5.521 & 4.424 & 0.969 & 0.887 & 5.943 & 4.673 & \textbf{0.967} & 0.865 & 7.226 & 5.772 & 0.938 & {0.806}
			\\
			\multicolumn{1}{c|}{}                                              & Spikformer \cite{zhou2022spikformer}                    
			& 9.988 & 7.875 & 0.800 & 0.630 & 9.948 & 7.851 & 0.796 & 0.619 & 9.964 & 7.988 & 0.821 & 0.638
			\\
			\multicolumn{1}{c|}{}                                              & TCN  \cite{bai2018empirical}                   
			& 4.523 & 3.399 & 0.962 & 0.920 & 4.487 & 3.438 & 0.961 & 0.921 & 6.662 & 4.794 & 0.923 & 0.837
			\\
			\multicolumn{1}{c|}{}                                              & LSTM \cite{hochreiter1997long}                    
			& 7.618 & 6.261 & 0.924 & 0.742 & 7.190 & 5.818 & 0.925 & 0.784 & 10.169 & 8.150 & 0.840 & 0.602

            \\
			\multicolumn{1}{c|}{}                                              & LS-SVM                    
			& 6.513 & 4.308 & 0.928 & 0.843 & 6.691 & 4.550 & 0.924 & 0.830 & 8.699 & 5.805 & 0.877 & 0.720
			
			\\  \hline   \hline

			\multicolumn{1}{c|}{\multirow{8}{*}{DB2}}                                    &
			\multicolumn{1}{c|}{\multirow{2}{*}{Method}}                                                                      & 
			\multicolumn{4}{c|}{Speed = 2.5}                                                                         & \multicolumn{4}{c|}{Speed = 3.0}                                                                                            
			& 
			\multicolumn{4}{c}{Speed = 3.6}
			\\ \cline{3-14}
			\multicolumn{1}{c|}{}      &\multicolumn{1}{c|}{}                                                     & \multicolumn{1}{c}{RMSE}                 & \multicolumn{1}{c}{MAE}          &\multicolumn{1}{c}{PCC}             &\multicolumn{1}{c|}{R\textsuperscript{2}}        
			& \multicolumn{1}{c}{RMSE}               & \multicolumn{1}{c}{MAE}            & \multicolumn{1}{c}{PCC}              &\multicolumn{1}{c|}{R\textsuperscript{2}}
			& \multicolumn{1}{c}{RMSE}               & \multicolumn{1}{c}{MAE}            & \multicolumn{1}{c}{PCC}               &\multicolumn{1}{c}{R\textsuperscript{2}}      \\ \cline{2-14}
			
			\multicolumn{1}{c|}{}                   & {SSAE}  
			& \textbf{3.632} & \textbf{2.971} & \textbf{0.945} & \textbf{0.877} & \textbf{3.388} & \textbf{2.720} & \textbf{0.939} & \textbf{0.835} & \textbf{2.969} & \textbf{2.348} & \textbf{0.928} & \textbf{0.853}
			\\ 
			\multicolumn{1}{c|}{}                                              & Informer  \cite{zhou2021informer}                
			& 4.136 & 3.421 & 0.930 & 0.838 & 4.931 & 4.025 & 0.922 & 0.749 & 4.451 & 3.655 & 0.914 & 0.795
			\\
			\multicolumn{1}{c|}{}                                              & Spikformer   \cite{zhou2022spikformer}                  
			& 8.581 & 7.145 & 0.694 & 0.448   & 8.898 & 7.279 & 0.664 & 0.411 & 8.333 & 6.563 & 0.721 & 0.493
			\\
			\multicolumn{1}{c|}{}                                              & TCN \cite{bai2018empirical}                     
			& 4.314 & 3.473 & 0.905 & 0.804 & 3.990 & 3.154 & 0.906 & 0.777 & 3.532 & 2.745 & 0.904 & 0.814
			\\
			\multicolumn{1}{c|}{}                                              & LSTM \cite{hochreiter1997long}                    
			& 6.133 & 4.867 & 0.750 & 0.583 & 6.111 & 4.952 & 0.712 & 0.531 & 6.453 & 5.336 & 0.673 & 0.534

            \\
			\multicolumn{1}{c|}{}                                              & LS-SVM                     
			& 5.596 & 4.803 & 0.896 & 0.485 & 5.181 & 4.392 & 0.866 & 0.541 & 5.456 & 4.736 & 0.853 & 0.623
			
			\\  \hline   \hline


	\end{tabular}}
	\label{tab2}
\end{table*}

\section{Experiments}\label{sec3}

We validated the effectiveness of SAFE-Net under two datasets, SIAT-DB1 (DB1) \cite{zhou2023continuous} and SIAT-DB2 (DB2) \cite{wang2022prediction}, representing different motion patterns. Before data collection, all volunteers provided formal written informed consent, and all experimental procedures were approved by the Medical Ethics Committee of the Shenzhen Institute of Advanced Technology ((SIAT)-IRB-200715-H0512).

\subsection{Datasets}
We validated the effectiveness of SAFE-Net under two datasets, DB1 and DB2, representing different motion patterns.

\subsubsection{DB1}

This dataset was collected from 8 healthy volunteers (age 23-26 years; weight 55-76 kg; height 170-182 cm). All individuals walked steadily on a treadmill for one minute at a speed of 3km/h at an incline of 0, 5, and 10.  We used Biometrics DataLOG (PS850) to collect sEMG signals from the subjects' gastrocnemius medialis (GM), rectus femoris (RF), vastus medialis (VM), vastus lateralis (VL), and biceps femoris (BF) muscles, with a sampling frequency of 500 Hz. Inertial measurement units (LPMS-B2) operating at 100 Hz were used to collect hip, knee, and ankle joint angles. During breaks between experiments, all volunteers were given sufficient rest to prevent muscle fatigue, and sensors were worn comfortably without restricting the volunteers' movements.

\subsubsection{DB2}
This dataset was collected from 12 healthy volunteers (age: 23-27 years, weight: 59.7-74.3 kg; Height: 164.3-174.7 cm). All individuals walked steadily on a horizontal treadmill for one minute at speeds of 2.5, 3.0, and 3.6 km/h, respectively. We used a 500 Hz sEMG device (TRIGNO, Delsys, Natick, MA, USA) to collect sEMG signals from the subjects' vastus medialis, rectus femoris, semitendinosus, tensor fasciae latae, peroneus longus, tibialis anterior, gastrocnemius medialis, and gastrocnemius lateralis muscles. A motion capture system (EAGLE, Motion Analysis, Rohnert Park, CA, USA) operating at 60 Hz was used to collect hip, knee, and ankle joint angles.

\begin{figure}[t]
	\centering
	\includegraphics[width=\linewidth]{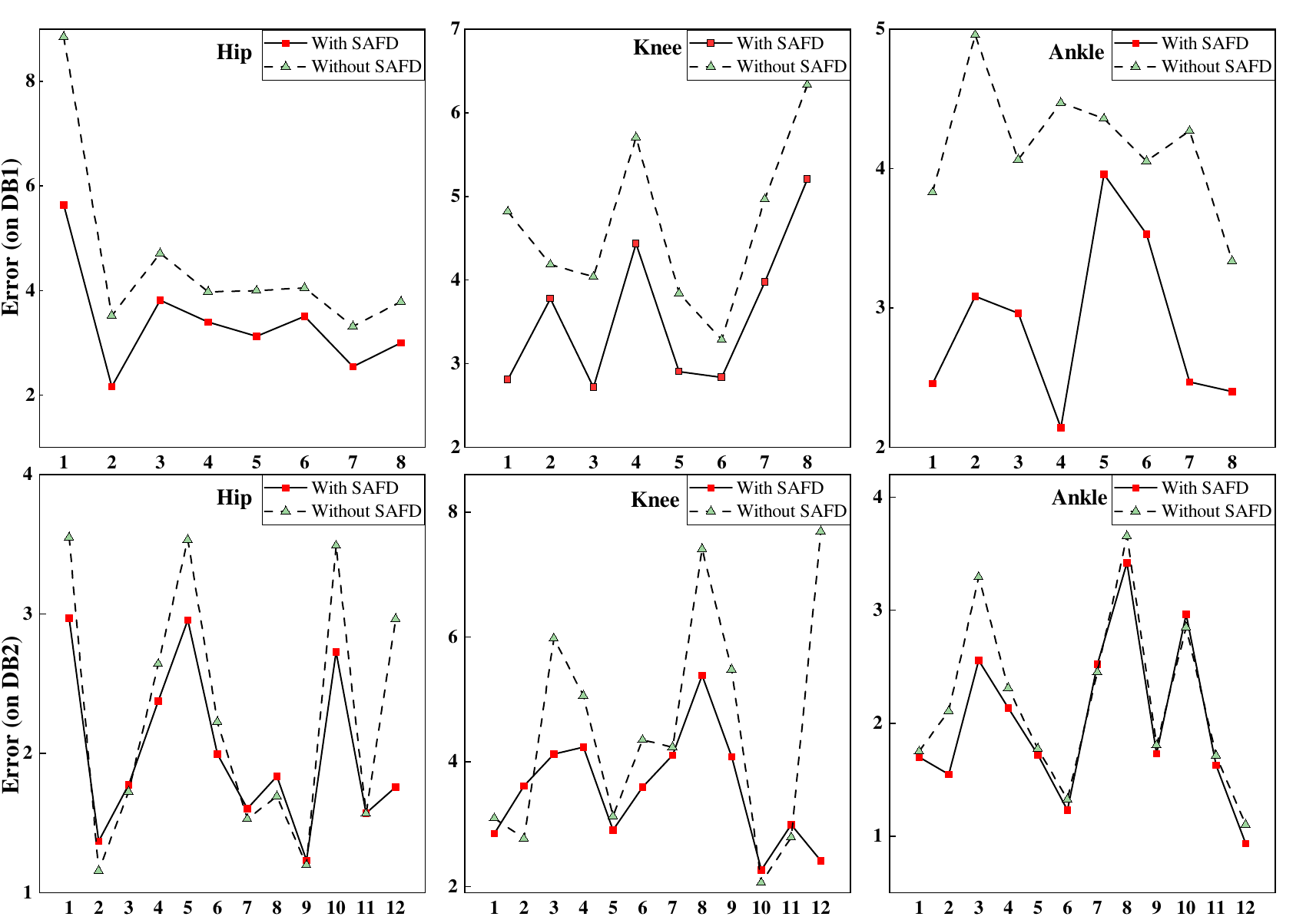}
	\caption{The line and symbol graphs present a comparison of results before and after applying SAFD on two datasets. The red squares represent SAFE-Net with SAFD, while the green triangles represent SSAE without SAFD. The horizontal and vertical axes denote the subject ID and RMSE, respectively.}
	\label{fig.SAFD}
\end{figure}

\subsection{Data preprocessing}
In DB1, the industrial frequency noise in the sEMG signals was removed using a second-order 50 Hz notch filter, followed by a fourth-order 20 Hz Butterworth high-pass filter. In DB2, the sEMG signals is removed from industrial noise using the same filter as in DB1, followed by a fourth-order 10 Hz Butterworth high-pass filter. To match the sEMG signals, the joint angle signals in both datasets were resampled to 500 Hz. Finally, all signals were standardized using the Z-score method, and segmented using an staggered sliding window approach with a window size of 100 ms and a step size of 16 ms. 

\subsection{Training protocol}
In the experiments, we partitioned the data of each individual into training, validation, and test sets in a ratio of 3:1:1. During training, we used a batch size of 50, iterated over all training data for 6 epochs, and employed an early stopping strategy. An adaptive Adam optimizer with an initial learning rate of 1e-4 was utilized in our experiments, with the learning rate adjusting adaptively as the training loss changes. On the test set, we evaluated the model's recognition accuracy and power consumption using  root-mean-squared error (RMSE) and theoretical power consumption \cite{xu2023novel}, respectively. Other hyperparameters are shown in Table \ref{tab1}. All experiments were conducted on an Intel Xeon Silver 4210R CPU @ 2.40GHz ($\times$2) and a NVIDIA RTX A6000 GPU.

\begin{table}[ht]
	\centering
	\caption{Comparison of inference costs between different models on DB1.}
	\begin{tabular}{p{1.2cm}p{1cm}p{1cm}p{1cm}p{1cm}p{1cm}}
		\toprule
		Model Name & Flops (M)  & Time (ms) & Power (W)  & Parameters (K) &Model Size (M)  \\
		\hline
		{SSAE} &\uline{67.68}   &\uline{0.56} &\uline{92.66} &\uline{27.14} & \uline{1.34}\\ 
		{Informer} &172.64     &0.87 &152.13 &69.32 & 2.74\\ 
		{Spikformer} &177.12   &0.92 &148.36 &70.08 & 2.77 \\
		{TCN} &\textbf{1.50}   &0.59 &\textbf{1.95} &\textbf{0.60} & 13.80 \\
		{LSTM} &133.60   &\textbf{0.35} &292.65 &52.42 & \textbf{0.20} \\
		\bottomrule
	\end{tabular}
	\label{tab3}
\end{table}

	%
%

\section{Results}\label{sec4}

\subsection{Comparison with different methods}

We mainly compared our proposed SSAE with two Transformer-based approaches, namely Informer \cite{zhou2021informer} and Spikformer \cite{zhou2022spikformer}. The comparison was conducted in terms of root mean squard error (RMSE), mean absolute error (MAE), Pearson correlation coefficient (PCC), and R-Square (R\textsuperscript{2}) for joint angle recognition, as shown in Table \ref{tab2}. When the slopes are 0°, 5°, and 10° respectively, SSAE reduced the average RMSE for hip-knee-ankle by 27.3\%, 28.0\%, and 19.8\% compared to Informer, and by 59.8\%, 57.0\%, and 41.9\% compared to Spikformer. When walking speeds are 2.5, 3.0, and 3.6 km/h, SSAE is 12.2\%, 31.3\%, and 33.3\% lower than the Informer hip-knee-ankle RMSE, and 57.7\%, 61.9\%, and 64.4\% lower than the Spikformer. It is worth noting that as the slope increases, the recognition error of Informer and Spikformer also increases, but the increase is the least for SSAE. Additionally, regardless of slope or speed, knee joint prediction is the most challenging among all joints, with SSAE achieving a maximum recognition RMSE of only 6.19 (reached at 2.5 km/h). This indicates that SSAE also exhibits good robustness under complex motion patterns.

Furthermore, we compared SSAE, Informer, and Spikformer on DB1 in terms of different evaluation metrics for inference costs: floating point operations ($Flops$), per-sample inference latency, theoretical power consumption, parameter count and model size. The theoretical power consumption was calculated according to the formula \cite{xu2023novel}, as follows:

\begin{equation}
	\label{eq.14}
	P=\frac{4.6 * \mathbf{\textit{MAC}}}{T}, \\
\end{equation}

\noindent where $T$ represents the total inference latency, and $MAC$ represents multiply-accumulate operations. In our experiments, we used $Flops$ instead of $MAC$. The results are shown in Table \ref{tab3}, where compared to Informer, SSAE reduced by 60.8\%, 34.9\%, 39.1\%, 60.8\%, and 51.09\% across the five metrics. Compared to Spikformer, SSAE reduced by 61.8\%, 38.3\%, 37.5\%, 61.3\%, and 51.62\%. SSAE exhibits lower inference costs compared to the other two methods. 

We also evaluated the performance of TCN \cite{bai2018empirical} and LSTM \cite{hochreiter1997long} in terms of recognition accuracy and inference costs. They both have a fatal drawback in that the model size of TCN and the power consumption of LSTM are several times higher than other models. Since least-squares support vector machine (LS-SVM) requires separate predictions for each joint angle and does not support parallel computation, we were unable to evaluate its inference cost. After weighing the trade-offs between prediction accuracy and inference costs, models based on the Transformer architecture are more suitable for SPC.

Finally, we visualized the RMSE of the hip-knee-ankle for each individual's data in both datasets, comparing SAFE-Net, Informer, and Spikformer. As shown in Fig. \ref{fig.diff_methods}, both on DB1 and DB2, our proposed SAFE-Net exhibits the lowest RMSE across all joints. On each individual's data, it can be seen that SAFE-Net achieves lower variance due to the fact that all the small triangles are more densely distributed and closer to the mean and median. Overall, SAFE-Net demonstrates superior generalization capabilities in multi-subject scenarios.


\subsection{Influence of the SAFD module}

We evaluated the performance of the SSAE and SAFE-Net structures before and after the inclusion of the SAFD module on two datasets. The results, depicted in Fig. \ref{fig.SAFD}, indicate that SAFE-Net with the SAFD module performs significantly better on DB1 compared to the SSAE without SAFD. On DB2, SAFE-Net shows advantages in hip-knee joints, while its performance in ankle joints is comparable to SSAE. Specifically, the average RMSE for the hip, knee, and ankle joints with and without SAFD are 2.01 vs. 2.27, 3.54 vs. 4.50, and 2.01 vs. 2.18, respectively. We evaluated the statistical significance of SSAE and SAFE-Net on ankle joint prediction separately in DB1 and DB2 datasets. One-way ANOVA was employed to compare the significant differences before and after the use of SAFD. In the case of DB1, the $P_{value}$ (1.93e-4) was found to be less than the significance level (0.05), and the $F_{test}$ statistic (16.43) was greater than the threshold value ($F_{crit}$ = 4.05), indicating a significant difference before and after the use of SAFD. Similarly, the same pattern was observed in DB2. From the perspective of different subject data, SAFE-Net indeed achieved lower recognition errors in multi-subject scenarios. This also demonstrates that the SAFD module can enhance the model's generalization ability.

\begin{figure}[t]
	\centering
	\includegraphics[height=2.7cm,width=0.35\linewidth]{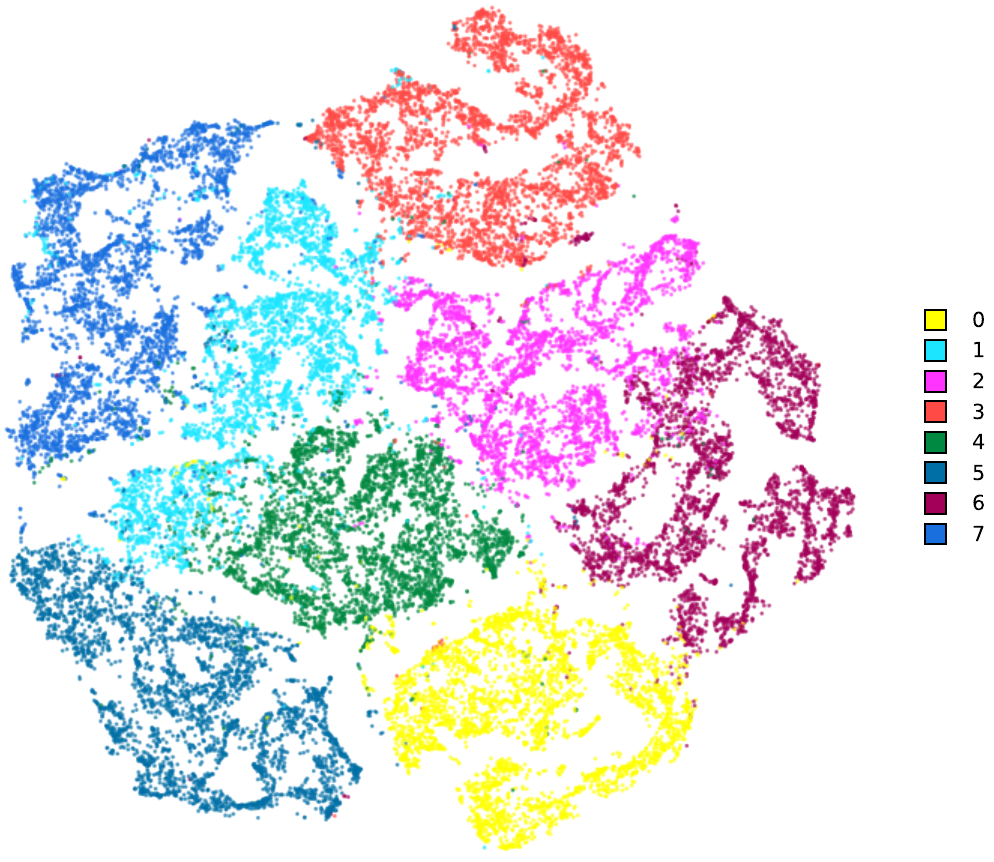}
	\caption{In DB1, the T-SNE visualization of the biological feature representations extracted by SAFE-Net is presented. The dataset includes data from eight subjects, with each color representing samples from different subjects.}
	\label{fig.cls}
\end{figure}

\section{Discussion}\label{sec5}

SSAE and SAFE-Net are proposed for estimating lower limb joint angles from sEMG and compared with Informer and Spikformer. RMSE and power consumption are used to measure recognition accuracy and inference cost, respectively. Experimental results show that, on both datasets, SSAE is significantly more efficient in the inference process, while SAFE-Net outperformed other models in multi-subject scenarios.

The SAFE-Net architecture addresses key limitations of physically based neuromuscular models, such as Hill-type muscle models, by leveraging data-driven and biologically inspired approaches. While Hill-type models are grounded in predefined biomechanical equations and require subject-specific parameters such as optimal muscle length and tendon stiffness. SAFE-Net eliminates the need for these constraints by learning directly from sEMG data. SAFE-Net advances beyond the rigid assumptions of Hill-type models, providing a scalable and efficient framework for tasks like multi-subject joint angle estimation, which is crucial for real-world SPC applications in rehabilitation robotics and wearable exoskeletons.

Based on spike-based attention, SAFE-Net is more akin to the operation mechanism of neurons, offering better biological plausibility. SNNs benefit from their unique spike transmission mechanism, improving the efficiency of the model's inference. The attention mechanism focuses only on the relevant positions and can model long-range temporal dependencies, serving as a biomimetic algorithm. Building a bridge between the two can provide good interpretability for decomposing sEMG signals. Specifically, in our study, we decomposed sEMG signals into kinematic and biological features, using them for angle regression and subject identification, respectively. In the previous experiments, we validated the effectiveness of kinematic features in angle regression tasks. 

To further demonstrate the decomposition of the sEMG signal into these two features is feasible, we used the t-stochastic neighbor embedding (T-SNE) \cite{van2008visualizing} visualization tool to observe the performance of SAFE-Net in subject identification. As shown in Fig. \ref{fig.cls}, we visualized the biological feature representation $F_b$ extracted by SAFE-Net in DB1. This dataset contains sEMG data from eight subjects, with each color representing a different subject sample. The figure shows that the $F_b$ features are divided into eight distinct colored regions, with clear boundaries between them and dense sample distributions within regions. This indicates that SAFE-Net is capable of distinguishing data from different subjects based on the extracted features. Therefore, SAFE-Net exhibits fault-tolerant and can handle partial loss of information through spiking sparsity and temporal coding. This makes it robust in handling data in multi-subject scenarios.


In fact, the theoretical power consumption of SSAE calculated according to Equation \ref{eq.14} is much higher than the actual value. In our experiments, we use $Flpos$ instead of $MAC$, and SSAE involves a large number of addition operations rather than multiplication operations, leading us to overestimate the power consumption of SSAE. Therefore, the proposed SSA mechanism has potential advantages in some energy-constrained applications. Overall, the promising SSAE and SAFE-Net in improving computational efficiency and muscle pattern recognition accuracy across subject scenarios are encouraging.

There are still some limitations to our work. The experimental data are collected under cyclic movement patterns, whereas human movement is characterized by complexity and variability, posing challenges for SAFE-Net in mixed movement scenarios. Moreover, our analysis exclusively focuses on data from healthy subjects, while individuals with walking disorders exhibit significantly distinct gait patterns compared to healthy individuals, necessitating further validation of SAFE-Net's generalizability. Expanding SAFE-Net to predict outcomes under mixed movement patterns and incorporating data from patients presents an intriguing avenue for future research.

For future work, we intend to gather data from diverse groups and movement patterns to enhance the validation of SAFE-Net's effectiveness. Further elucidation of the physiological and kinematic mechanisms underlying the decomposition of sEMG signals is warranted. Drawing inspiration from expression recognition, the SAFD strategy proposed in this study offers a framework for decomposing sEMG into distinct kinematic and biological features, thereby paving the way for delving into biological interpretability. Additionally, while spiking neural networks play a pivotal role in enhancing computational efficiency, their full potential remains untapped and warrants comprehensive exploration in future research endeavors.

\section{Conclusion}\label{sec6}
In this study, we introduced a novel SAFE-Net neural architecture for continuous joint angle estimation. The architecture comprises two components, SSAE and SAFD which are designed to reduce inference costs and enhance recognition accuracy in multi-subject scenarios, respectively. SAFE-Net demonstrated state-of-the-art performance on two datasets. Moreover, the model's inference costs are much lower than that of Informer and Spikformer. Overall, we have developed an efficient sEMG encoder and hierarchical feature decoder for joint angle prediction tasks. The proposed SAFE-Net represents a significant advancement towards the realization of lower limb rehabilitation exoskeletons with synchronous and proportional control, and can also be extended for other myoelectric pattern recognition tasks.


\bibliographystyle{IEEEtran}
\normalem
\bibliography{root}{}

\end{document}